\documentclass[sigconf,screen]{acmart}
\usepackage{float}  

\usepackage{multirow}
\usepackage{tikz}
\usetikzlibrary{shapes.geometric, arrows.meta, positioning, calc, decorations.pathreplacing}
\usepackage{enumitem}

\AtBeginDocument{%
  }

\usepackage{xcolor}

\setcopyright{cc}
\setcctype{by}
\acmDOI{10.1145/3832783.3837546}
\acmYear{2026}
\copyrightyear{2026}
\acmISBN{979-8-4007-2882-2/2026/10}
\acmConference[ASE '26]{Proceedings of the 41st IEEE/ACM International Conference on Automated Software Engineering}{October 12--16, 2026}{Munich, Germany}
\acmBooktitle{Proceedings of the 41st IEEE/ACM International Conference on Automated Software Engineering (ASE '26), October 12--16, 2026, Munich, Germany}
\acmSubmissionID{ase26main-p2970-p}
\received{2026-03-26}
\received[accepted]{2026-06-18}

\begin{document}

\title{A Few Pages of Markdown: Committed AI Configuration and Lower Quality Cost after Coding-Agent Adoption}

\author{Yegor Denisov-Blanch}
\orcid{0009-0009-0967-8399}
\authornote{Joint first authors, contributed equally; author ordering decided by token spend.}
\affiliation{%
  \institution{Stanford University}
  \city{Stanford}
  \country{USA}
}

\author{Shyam Agarwal}
\correspondingauthor
\orcid{0009-0009-2147-5674}
\authornotemark[1]
\affiliation{%
  \institution{Carnegie Mellon University}
  \city{Pittsburgh}
  \country{USA}
}

\author{Pavel Azaletskiy}
\orcid{0000-0003-3884-4324}
\authornotemark[1]
\affiliation{%
  \institution{Grid Dynamics}
  \city{Cleveland}
  \country{USA}
}

\author{Hao He}
\orcid{0000-0001-8311-6559}
\affiliation{%
  \institution{Carnegie Mellon University}
  \city{Pittsburgh}
  \country{USA}
}

\author{Rylan Schaeffer}
\orcid{0000-0002-4298-7216}
\affiliation{%
  \institution{Stanford University}
  \city{Stanford}
  \country{USA}
}

\author{Brando Miranda}
\orcid{0009-0008-5031-0126}
\affiliation{%
  \institution{Stanford University}
  \city{Stanford}
  \country{USA}
}

\author{Bogdan Vasilescu}
\orcid{0000-0003-4418-5783}
\affiliation{%
  \institution{Carnegie Mellon University}
  \city{Pittsburgh}
  \country{USA}
}

\author{Sanmi Koyejo}
\orcid{0000-0002-4023-419X}
\affiliation{%
  \institution{Stanford University}
  \city{Stanford}
  \country{USA}
}

\renewcommand{\shortauthors}{Denisov-Blanch, Agarwal, Azaletskiy et al.}

\begin{abstract}
Coding agents increase development velocity but also technical debt.
Prior work reports only average effects across adopters, hiding wide differences between teams.
We introduce RAMP (Repository AI Maturity Profile), a four-level cumulative maturity model grounded in version-controlled artifacts that teams commit to configure AI tools.
RAMP runs from behavioral rules and coding standards through named agent definitions to multi-agent orchestration, with observed practice concentrated in the first three levels.
Across 441 repositories the levels behave as a cumulative scale, and independent human annotation reproduces RAMP's repository-level labels on 97\% of a held-out sample.
Adoption is cumulative, forward-only, and set-and-forget: 73.8\% of artifacts are committed once and never modified.
Re-estimating an existing agent-adoption panel within each stratum, agents accelerate development regardless of maturity (28--38\% more commits), but quality diverges: among agent-first repositories, where the contrast is identified, those without committed AI configuration show roughly twice the increase in cognitive complexity (+53\% versus +27\%) and 1.7× the increase in static-analysis warnings.
Because maturity is observational, correlated engineering discipline or model capability may explain part of the gap; we present these findings as hypothesis-generating and release RAMP as a reusable instrument.

\end{abstract}



\maketitle

\section{Introduction}

Large language model (LLM)-powered coding assistants have moved from novelty to near-ubiquity in under three years, with a majority of professional developers now using AI assistance during daily work~\cite{stackoverflow2024, github2024survey}.
A new generation of autonomous coding agents extends this trend further, generating entire pull requests with minimal human intervention.
Teams report sharply different returns: some prominent developers report large, durable gains from agentic workflows~\cite{ronacher2025, willison2026}, while a countercurrent on developer forums describes AI output as slop that burdens reviewers and maintainers~\cite{baltes2026slop}.
The empirical evidence splits along similar lines, with controlled experiments finding short-term productivity gains from AI assistance~\cite{peng2023, ziegler2024} and longitudinal studies reporting persistent increases in static-analysis warnings, code complexity, and maintenance burden~\cite{he2026cursor, agarwal2026, xu2025, gitclear2025}.
If both accounts are accurate, what separates the teams these tools help from those they hurt is an open question.
Existing studies cannot answer this: they report only average effects across adopters, and none examine whether \textit{how} a repository structures its AI integration is associated with different outcomes.

Teams differ widely in how they set up AI tools, and part of that setup is committed to version control, where it becomes a concrete, observable signal.
The variation is also ordered, from no shared configuration toward increasingly deliberate structure.
Software engineering has long captured progressions of this kind with maturity models~\cite{cmmi, herbsleb1997, dora2024, forsgren2018}, and the appetite for an AI-specific one is clear: the Software Engineering Institute and several large consultancies have each proposed one in the past year~\cite{sei2026, deloitte2026, kpmg2025, pwc2026}.
Such a model is what our question calls for, but these frameworks score an organization through leadership surveys and self-assessment, which cannot connect a repository's committed AI configuration to its code-quality outcomes.
Quality outcomes are properties of a codebase, so practice must be measured at the same unit; and many adoption events have already happened, so the measurement must come from what teams left behind.

This paper builds that instrument and then uses it.
We formalize the progression as RAMP (Repository AI Maturity Profile), a four-level cumulative model grounded in the artifacts teams commit to version control: repositories with no committed configuration sit at Level~1; files that supply project context, such as coding conventions and behavioral rules, mark Level~2; reusable capabilities, such as named agents and packaged commands, mark Level~3; and multi-step workflows that coordinate several agents mark Level~4.
The progression runs from context to capabilities to coordination.
To operationalize the model, we develop a classification pipeline that combines filename-pattern heuristics with embedding-based semantic classification, covering 12 AI coding tools across 1,046 validated artifacts in 441 repositories. Guttman scalogram analysis confirms a cumulative hierarchy over the observed levels, and the classifier reproduces human-derived maturity levels for 34 of 35 repositories in a held-out annotation study (Section~\ref{sec:ramp}).

We deploy the validated classifier in two complementary studies (Figure~\ref{fig:study-design}).
Study~1 characterizes the landscape and dynamics of AI practice adoption across corporate repositories, finding that adoption is cumulative, forward-only, and predominantly set-and-forget.
Maturity is therefore a stable property of a repository, which makes it usable as a stratification variable in Study~2, where we reanalyze the agent-adoption panel of Agarwal et al.~\cite{agarwal2026}, augmenting each repository with its RAMP level to test whether practice maturity moderates the effects of coding agent adoption.
We find that agents accelerate development across maturity strata, but quality degradation concentrates in repositories without committed AI configuration: in the agent-first stratum, where the maturity contrast is identified, complexity increases roughly 2$\times$ more at Level~1 than at Level~2 and above.
Because maturity is observational, we treat these associations as hypothesis-generating and discuss alternative explanations in Section~\ref{sec:threats}.

Overall, this paper makes the following contributions:

\begin{enumerate}

\item RAMP, a four-level cumulative maturity model for AI-assisted development grounded in version-controlled artifacts.

\item A classification pipeline that operationalizes the model across 12 AI coding tools using an ensemble of tool-pattern matching, path-semantic embedding, and content-semantic embedding, validated against human annotation.

\item An empirical characterization of AI practice maturity across 441 corporate repositories, showing that adoption is cumulative, forward-only, and predominantly set-and-forget.

\item Exploratory evidence that practice maturity is associated with quality outcomes after coding agent adoption.
\end{enumerate}

\section{Background and Related Work}
\label{sec:background}

\subsection{AI Coding: Velocity Gains and Quality Costs}

Studies of IDE-based AI assistants report mostly consistent short-term velocity gains. Randomized and quasi-experimental evaluations of GitHub Copilot find faster task completion and higher code throughput, with magnitudes varying by developer experience and task complexity~\cite{peng2023, ziegler2024, paradis2024, imai2022}. Enterprise trials confirm the direction of these effects~\cite{chretien2024}. Quality findings are less settled. Some controlled studies find no significant increase in security vulnerabilities from AI-assisted code~\cite{asare2022}. He et al.~\cite{he2026cursor} provide the most detailed longitudinal evidence for IDE-based tools, studying Cursor adoption across open-source projects; they find large but transient velocity gains alongside persistent increases in static-analysis warnings (30\%) and cognitive complexity (41\%), with quality degradation acting as a drag on future velocity. Xu et al.~\cite{xu2025} show that AI-assisted code requires more rework to meet repository standards, with the maintenance burden falling disproportionately on experienced core developers. The pattern across this literature is that velocity effects are reliably positive in the short term, while quality effects depend on the time horizon, the metric, and the setting.

Work on autonomous coding agents is newer and the evidence base thinner. Watanabe et al.~\cite{watanabe2025} find that maintainers merge 84\% of agent-generated pull requests, indicating high acceptance, though their study examines merge behavior and not downstream code quality. The most directly relevant predecessor is Agarwal et al.~\cite{agarwal2026}, who use staggered difference-in-differences to estimate causal effects of agent adoption across open-source repositories. They report large velocity gains in repositories where agents are the first AI tool and minimal gains where IDE-based tools preceded agents. Quality degradation (roughly 18\% more warnings, 39\% higher complexity) persists across both settings. Their agent-first versus IDE-first partition captures a binary measure of prior AI exposure but does not capture how repositories \textit{structure} their AI integration. Two repositories that both adopted agents without prior IDE experience may differ substantially in whether they have committed behavioral rules, configuration files, or workflow definitions that constrain how agents operate. RAMP captures this dimension: the committed artifacts that shape agent behavior within a repository.

All studies in this literature report average effects across adopters. None examine whether the presence of committed AI configuration artifacts in a repository is associated with different outcomes. He et al.'s effect sizes (30\% warning increase, 41\% complexity increase) provide a useful benchmark: if practice maturity matters, repositories with structured practices should show effects below these averages, and repositories without them should show effects above.

\subsection{Process Maturity in Software Engineering}

The idea that organizational maturity conditions technology effectiveness has a long history in software engineering. The Capability Maturity Model (CMM) formalized process maturity into staged levels, and empirical work linked higher CMM levels to lower defect rates, better schedule adherence, and improved project outcomes~\cite{cmmi, herbsleb1997}. CMMI extended this framework across disciplines. More recently, the DORA program's State of DevOps research has demonstrated that specific engineering practices (continuous integration, trunk-based development, monitoring) predict delivery performance and stability more reliably than the adoption of any particular tool or platform~\cite{dora2024, forsgren2018}. Across three decades of this research, the recurring finding is that the maturity of a team's practices predicts outcomes more reliably than the tools the team uses.
These models rely on organizational self-assessment or developer surveys, which are poorly suited to AI tool integration, where practices are new, terminology is unstandardized, and teams may not distinguish between ad hoc prompting and structured configuration.
Version-controlled artifacts avoid these problems: they are objective, contemporaneous, and extractable at scale.

AI coding tools raise an analogous question. Teams vary widely in how they configure AI tools in version control, from committing nothing at all to maintaining orchestration workflows that coordinate multi-agent tasks. These committed artifacts represent deliberate, version-controlled choices about how AI tools should operate in a codebase, visible to all contributors. A rules file that constrains agent behavior, for example, shapes every AI-generated contribution to the repository, making these artifacts a plausible channel through which team-level practices could influence code quality outcomes. Whether these artifacts form a structured hierarchy, and whether that structure is associated with different development outcomes after agent adoption, are open empirical questions that this paper addresses.

\subsection{Research Questions and Study Overview}

The literature reviewed above motivates two research questions corresponding to two complementary studies.

Study~1 characterizes the dynamics of AI practice adoption across repositories, using the RAMP maturity classification developed and validated in Section~\ref{sec:ramp}.

\textbf{RQ1} (Adoption dynamics): \textit{How do repositories transition between maturity levels over time, and what proportion of transitions are forward, stalled, or reversed?} Understanding whether maturity levels reflect a stable property of repositories or a fragile state that commonly reverses determines whether they are useful as a stratification variable in downstream studies.

Study~2 applies the classification to an existing panel dataset to explore whether practice maturity is associated with heterogeneous effects after coding agent adoption.

\textbf{RQ2} (Outcomes): \textit{Is AI practice maturity associated with differential outcomes after coding agent adoption, and does the association differ between quality metrics and velocity metrics?} Prior work establishes that agent adoption increases code complexity and static-analysis warnings on average~\cite{agarwal2026}. RQ2 asks whether that average effect conceals heterogeneity linked to practice maturity.

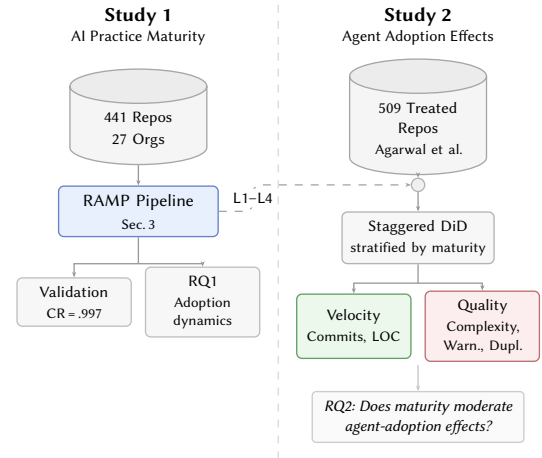
\begin{figure}[!t]
\centering
\vspace{0.5em}%
\begin{tikzpicture}[
    font=\sffamily\fontsize{6.5}{8}\selectfont,
    >={Stealth[length=2pt, width=1.5pt]},
    line width=0.5pt,
    databox/.style={
        cylinder,
        cylinder uses custom fill,
        cylinder body fill={gray!8},
        cylinder end fill={gray!15},
        draw=gray!70,
        shape border rotate=90,
        minimum width=1.8cm,
        minimum height=0.8cm,
        text width=1.6cm,
        align=center,
        aspect=0.25,
        font=\sffamily\fontsize{6}{7.5}\selectfont,
    },
    procbox/.style={
        rectangle,
        rounded corners=2pt,
        draw=gray!60,
        fill={gray!6},
        minimum height=0.6cm,
        text width=1.8cm,
        align=center,
        font=\sffamily\fontsize{6}{7.5}\selectfont,
    },
    rampbox/.style={
        rectangle,
        rounded corners=2pt,
        draw={rgb,255:red,100;green,130;blue,200},
        fill={rgb,255:red,232;green,240;blue,254},
        minimum height=0.6cm,
        minimum width=2.1cm,
        text width=1.9cm,
        align=center,
        font=\sffamily\fontsize{6.5}{8}\selectfont,
    },
    rqbox/.style={
        rectangle,
        rounded corners=2pt,
        draw=gray!50,
        fill={gray!6},
        minimum height=0.6cm,
        text width=1.3cm,
        align=center,
        font=\sffamily\fontsize{6}{7.5}\selectfont,
    },
    outgreen/.style={
        rqbox,
        minimum height=0.85cm,
        draw={rgb,255:red,100;green,160;blue,100},
        fill={rgb,255:red,235;green,248;blue,235},
    },
    outred/.style={
        rqbox,
        minimum height=0.85cm,
        draw={rgb,255:red,190;green,110;blue,110},
        fill={rgb,255:red,252;green,238;blue,238},
    },
    arr/.style={
        draw=gray!80,
        -{Stealth[length=2pt, width=1.5pt]},
        line width=0.5pt,
    },
    darr/.style={
        draw=gray!70,
        -{Stealth[length=2pt, width=1.5pt]},
        line width=0.5pt,
        dashed,
    },
    panellabel/.style={
        font=\sffamily\bfseries\fontsize{7.5}{9}\selectfont,
    },
    sublabel/.style={
        font=\sffamily\fontsize{6}{7.5}\selectfont,
        text=black,
    },
]

\def\lx{-1.85}
\def\rx{1.85}
\def\divx{0}
\def\sp{0.85}

\draw[gray!35, dashed, line width=0.4pt]
    (\divx, 0.35) -- (\divx, -5.75);

\node[panellabel] at (\lx, 0.2) {Study 1};
\node[sublabel] at (\lx, -0.08) {AI Practice Maturity};

\node[databox] (repo) at (\lx, -1.3)
    {441 Repos\\[0.5pt]27 Orgs};

\node[rampbox] (ramp) at (\lx, -2.38)
    {RAMP Pipeline\\[0.5pt]
     {\fontsize{5.5}{6.5}\selectfont Sec.\,3}};

\node[rqbox] (val) at ($(\lx,0)+(-\sp,-3.58)$)
    {Validation\\[0.5pt]{\fontsize{5.5}{6.5}\selectfont CR\,=\,.997}};
\node[rqbox] (rq1) at ($(\lx,0)+(\sp,-3.58)$)
    {RQ1\\[0.5pt]{\fontsize{5.5}{6.5}\selectfont Adoption}\\[-0.5pt]
     {\fontsize{5.5}{6.5}\selectfont dynamics}};

\draw[arr] (repo) -- (ramp);
\draw[draw=gray!80, line width=0.5pt] (ramp.south) -- ++(0,-0.35) coordinate (fork1);
\draw[arr] (fork1) -| (val.north);
\draw[arr] (fork1) -| (rq1.north);

\node[panellabel] at (\rx, 0.2) {Study 2};
\node[sublabel] at (\rx, -0.08) {Agent Adoption Effects};

\node[databox] (treated) at (\rx, -1.3)
    {509 Treated Repos\\[0.5pt]{\fontsize{5.5}{6.5}\selectfont
     Agarwal et al.}};

\node[circle, draw=gray!60, fill=gray!12,
      inner sep=0pt, minimum size=5pt] (merge) at (\rx, -2.03) {};

\node[procbox] (did) at (\rx, -2.73)
    {Staggered DiD\\[0.5pt]
     {\fontsize{5.5}{6.5}\selectfont stratified by maturity}};

\node[outgreen] (vel) at ($(\rx,0)+(-\sp,-3.90)$)
    {Velocity\\[0.5pt]{\fontsize{5.5}{6.5}\selectfont Commits, LOC}};
\node[outred] (qual) at ($(\rx,0)+(\sp,-3.90)$)
    {Quality\\[0.5pt]{\fontsize{5.5}{6.5}\selectfont Complexity,}\\[-0.5pt]
     {\fontsize{5.5}{6.5}\selectfont Warn., Dupl.}};

\node[rectangle, rounded corners=1.5pt, draw=gray!40, fill=gray!4,
      minimum width=2.8cm, align=center,
      font=\sffamily\itshape\fontsize{6}{7.5}\selectfont]
      (rq2) at (\rx, -5.10)
    {RQ2: Does maturity moderate\\agent-adoption effects?};

\draw[arr] (treated) -- (merge);
\draw[arr] (merge) -- (did);
\draw[draw=gray!80, line width=0.5pt] (did.south) -- ++(0,-0.25) coordinate (fork2);
\draw[arr] (fork2) -| (vel.north);
\draw[arr] (fork2) -| (qual.north);
\draw[arr, gray!45]
    ($(vel.south)!0.5!(qual.south)+(0,-0.06)$) -- (rq2);

\draw[darr, rounded corners=3pt]
    (ramp.east) -- ++(0.45,0) |- (merge.west)
    node[pos=0.25, fill=white, inner sep=1.5pt,
         font=\sffamily\fontsize{6}{7.5}\selectfont]
    {L1--L4};

\end{tikzpicture}
\vspace{-0.3cm}
\caption{Study design. Study~1 develops and validates RAMP across 441 corporate repositories. Study~2 stratifies agent-adoption effects by RAMP maturity levels.}
\label{fig:study-design}
\Description{A two-panel flowchart. The left panel, Study~1, runs from a data cylinder labeled 441 repositories across 27 organizations into a RAMP pipeline box, which forks into a validation box reporting CR = 0.997 and an RQ1 box on adoption dynamics. The right panel, Study~2, runs from a cylinder of 509 treated repositories from Agarwal et al.\ into a staggered difference-in-differences box stratified by maturity, which forks into a green velocity box (commits, lines of code) and a red quality box (complexity, warnings, duplication); both feed a question box asking whether maturity moderates agent-adoption effects. A dashed arrow labeled L1--L4 carries maturity labels from the RAMP pipeline on the left into the Study~2 flow on the right.}
\end{figure}

The RAMP model is developed and validated on 441 corporate repositories from 27 organizations (Section~\ref{sec:ramp}). Study~1 analyzes adoption dynamics on the 196-repository temporal subset with reconstructed git histories. Study~2 applies the classification to the open-source analysis sample of Agarwal et al.~\cite{agarwal2026}. Of the 518 repositories that adopted AI coding agents in Study~2, 9 could not be traced to an accessible snapshot; the remainder were classified by running the RAMP pipeline on repositories not included in the development sample.\footnote{Guttman and related validation metrics for RAMP (Section~\ref{sec:validation}) are computed on the 441-repository development sample. Study~2 applies the same frozen classifier to Agarwal et al.'s repositories, which are disjoint from that development set. A direct human-annotation validation of the classifier, on a held-out sample drawn from the Study~2 corpus, is reported in Section~\ref{sec:human-validation}. Supplementary robustness checks are reported in the replication package.} Figure~\ref{fig:study-design} shows the overall design.

\section{RAMP: Repository AI Maturity Profile}
\label{sec:ramp}
\label{sec:ramp_data}

The RAMP classifier was developed on 441 private GitHub repositories from 27 commercial organizations,\footnote{This development sample is disjoint from the Study~2 dataset.} recruited through ongoing research on AI development practices and spanning diverse industry sectors, team sizes, and development cultures. Below we describe the classification pipeline and validate the resulting maturity scale.

\smallskip\noindent\textbf{Collection and filtering.}
A broad discovery sweep over these organizations' repositories returned 30,216 candidate artifacts across 1,931 repositories.
Four inclusion/exclusion filters---removing single-artifact repositories (too sparse to profile), boilerplate project files (e.g., \texttt{README}, \texttt{LICENSE}) swept in beyond tool-specific patterns, outlier repositories in the top 10\% by artifact count (template collections on inspection), and near-duplicate files detected during classification---yield the 441-repository development frame with roughly 4{,}300 candidate files.
Of these 441 repositories, 217 have at least one validated AI artifact (1{,}046 artifacts; 1{,}026 scoreable across 210 repositories after boilerplate and documentation-folder pre-filters)---these 217 form the \textit{AI-tools subset} used in robustness checks; 231 repositories retain no scoreable artifact and a further 63 of the 210 yield only non-leveled documentation, so 294 fall at Level~1 by construction (Section~\ref{sec:validation}).

\begin{table*}[t]
\caption{\textbf{RAMP: Repository AI Practice Maturity.}
Level = highest level where $\geq$1 artifact exists. Levels are cumulative; higher-level repositories typically contain lower-level artifacts as well.$^{\dagger}$}
\label{tab:levels}
\small
\setlength{\tabcolsep}{4pt}
\renewcommand{\arraystretch}{0.7}
\begin{tabular}{p{1.4cm}p{7.2cm}p{3.2cm}p{4cm}}
\toprule
\textbf{Level} & \textbf{What exists in the repo} {\footnotesize (any one is sufficient)} & \textbf{What this enables} & \textbf{What changed} \\

\midrule
\textbf{L1} \newline \textit{Unconfigured}
& No AI-related files committed.
& AI has no project knowledge; each session starts from scratch.
& ---
\\

\midrule
\textbf{L2} \newline \textit{Grounded} \newline \textit{Prompting}
& \textbf{Context artifacts:}
  \begin{itemize}[leftmargin=0.8cm, itemsep=0pt, parsep=0pt, topsep=0pt, after=\vspace{-0.1cm}]
  \item Behavioral rules or instructions (\textit{rules})
  \item AI tool settings (\textit{config})
  \item Architecture or design docs (\textit{architecture})
  \item Coding standards with examples (\textit{code-style})
  \end{itemize}
& AI is project-aware; outputs align with team conventions across all contributors.
& The AI now knows how your project works. L1 starts from a blank slate; L2 gives it shared context.
\\

\midrule
\textbf{L3} \newline \textit{Agent-} \newline \textit{Augmented}
& \textbf{Capability artifacts:}
  \begin{itemize}[leftmargin=0.8cm, itemsep=0pt, parsep=0pt, topsep=0pt, after=\vspace{-0.1cm}]
  \item Named agents with roles and tool restrictions (\textit{agents})
  \item Reusable commands or prompt templates (\textit{commands})
  \item Domain-knowledge guides (\textit{skills})
  \end{itemize}
& AI performs structured, repeatable tasks via specialized roles.
& L2 tells the AI how to \textit{behave} (``never use magic numbers''). L3 teaches it how to \textit{do things} (``here is how we do database migrations'').
\\

\midrule
\textbf{L4} \newline \textit{Orchestration}
& \textbf{Coordination artifacts:}
  \begin{itemize}[leftmargin=0.8cm, itemsep=0pt, parsep=0pt, topsep=0pt, after=\vspace{-0.1cm}]
  \item Multi-agent workflows with task assignments (\textit{flows})
  \item Pipelines with phases and dependencies
  \item Execution logs tracking agent actions (\textit{session-logs})
  \end{itemize}
& Multiple agents coordinated into end-to-end workflows with defined handoffs.
& L3 defines what each agent can do alone. L4 defines how agents pass work to one another.
\\
\bottomrule
\multicolumn{4}{@{}p{\dimexpr 1.4cm+7.2cm+3.2cm+3.3cm+6\tabcolsep\relax}@{}}{\footnotesize $^{\dagger}$\,Levels are tool-independent. A Cursor user with only \texttt{.cursorrules} is L2; the same tool with agents and workflows is L4. Level~4 is defined by the model but not observed in the development sample (Section~\ref{sec:validation}); it does appear in the open-source sample of Study~2.} \\

\end{tabular}
\end{table*}

\subsection{Classification Pipeline}
\label{sec:pipeline}

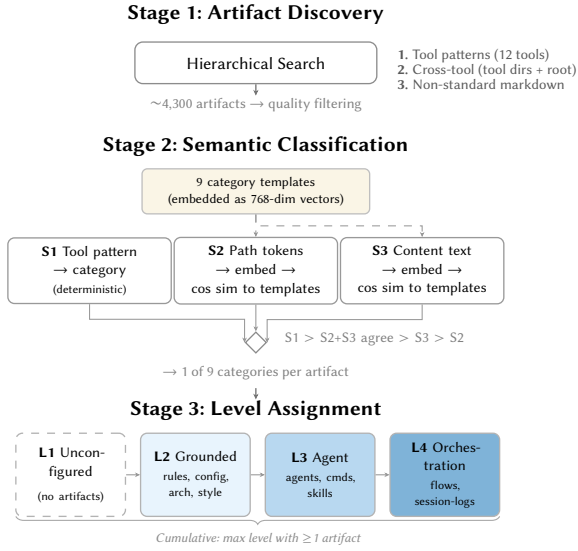
\begin{figure}[t]
\centering
\begin{tikzpicture}[
    font=\sffamily\fontsize{7}{8.5}\selectfont,
    >={Stealth[length=2.2pt, width=2pt]},
    line width=0.5pt,
    stagehead/.style={
        font=\sffamily\bfseries\fontsize{8}{9.5}\selectfont,
        fill=white, inner sep=2pt,
    },
    procbox/.style={
        rectangle, rounded corners=3pt,
        draw=black!45, fill=white,
        minimum height=0.55cm, align=center,
        font=\sffamily\fontsize{6.5}{8}\selectfont,
    },
    sigbox/.style={
        rectangle, rounded corners=2.5pt,
        draw=black!40, fill=white,
        minimum height=0.55cm, minimum width=2.05cm,
        text width=1.95cm, align=center,
        font=\sffamily\fontsize{5.5}{7}\selectfont,
    },
    templatebox/.style={
        rectangle, rounded corners=2pt,
        draw=black!25, fill={rgb,255:red,250;green,245;blue,230},
        minimum height=0.4cm, minimum width=3.0cm,
        text width=2.8cm, align=center,
        font=\sffamily\fontsize{5}{6.5}\selectfont,
    },
    stepitem/.style={
        font=\sffamily\fontsize{5.5}{7}\selectfont,
        text=black!65, anchor=base west,
    },
    annot/.style={
        font=\sffamily\fontsize{5.5}{7}\selectfont, text=black!50,
    },
    arr/.style={
        draw=black!55,
        -{Stealth[length=2.2pt, width=2pt]},
        line width=0.5pt,
    },
    arrthin/.style={
        draw=black!40,
        -{Stealth[length=1.8pt, width=1.6pt]},
        line width=0.4pt,
    },
    arrnone/.style={
        draw=black!40, line width=0.4pt,
    },
]

\node[stagehead] (s1head) at (0, 0) {Stage 1: Artifact Discovery};

\node[procbox, minimum width=3.2cm] (search) at (0, -0.65)
    {Hierarchical Search};

\foreach \i/\txt [count=\n] in {
    1/Tool patterns (12 tools),
    2/Cross-tool (tool dirs + root),
    3/Non-standard markdown%
} {
    \node[stepitem]
        at ($(search.east)+(0.15, 0.25 - \n*0.2)$)
        {\textbf{\i.}\;\txt};
}

\node[annot] at (0, -1.2)
    {{\footnotesize$\sim$}4{,}300 artifacts $\rightarrow$ quality filtering};

\node[stagehead] (s2head) at (0, -1.75) {Stage 2: Semantic Classification};

\node[templatebox] (templates) at (0, -2.35)
    {9 category templates\\(embedded as 768-dim vectors)};

\def\sigy{-3.4}
\node[sigbox] (sig1) at (-2.2, \sigy)
    {\textbf{S1} Tool pattern\\$\rightarrow$ category\\{\fontsize{4.5}{5.5}\selectfont(deterministic)}};
\node[sigbox] (sig2) at (0, \sigy)
    {\textbf{S2} Path tokens\\$\rightarrow$ embed $\rightarrow$\\cos sim to templates};
\node[sigbox] (sig3) at (2.2, \sigy)
    {\textbf{S3} Content text\\$\rightarrow$ embed $\rightarrow$\\cos sim to templates};

\draw[arrthin, dashed] (templates.south) -- ++(0,-0.15) -| (sig2.north);
\draw[arrthin, dashed] (templates.south) -- ++(0,-0.15) -| (sig3.north);

\node[diamond, draw=black!45, fill=white,
      inner sep=0pt, minimum size=8pt] (merge) at (0, -4.35) {};

\node[annot, text width=3.8cm, anchor=west, align=left] at ($(merge.east)+(0.12,0.05)$)
    {S1 $>$ S2{+}S3 agree $>$ S3 $>$ S2};

\node[annot] (catout) at (0, -4.75)
    {$\rightarrow$ 1 of 9 categories per artifact};

\draw[arrthin] (sig1.south) -- ++(0,-0.2) -| (merge.west);
\draw[arrthin] (sig2.south) -- (merge.north);
\draw[arrthin] (sig3.south) -- ++(0,-0.2) -| (merge.east);

\node[stagehead] (s3head) at (0, -5.25) {Stage 3: Level Assignment};

\def\ly{-6.1}

\node[rectangle, rounded corners=2.5pt,
      draw=black!30, dashed, fill=white,
      minimum width=1.45cm, minimum height=1.1cm,
      text width=1.2cm, align=center,
      font=\sffamily\fontsize{5.5}{7}\selectfont]
    (l1) at (-2.45, \ly)
    {\textbf{L1} Unconfigured\\[2pt]{\fontsize{4.5}{5.5}\selectfont(no artifacts)}};

\node[rectangle, rounded corners=2.5pt,
      draw=black!30,
      fill={rgb,255:red,232;green,244;blue,253},
      minimum width=1.45cm, minimum height=1.1cm,
      text width=1.2cm, align=center,
      font=\sffamily\fontsize{5.5}{7}\selectfont]
    (l2) at (-0.8, \ly)
    {\textbf{L2} Grounded\\{\fontsize{4.5}{5.5}\selectfont rules, config,}\\[-1pt]{\fontsize{4.5}{5.5}\selectfont arch, style}};

\node[rectangle, rounded corners=2.5pt,
      draw=black!30,
      fill={rgb,255:red,184;green,216;blue,240},
      minimum width=1.45cm, minimum height=1.1cm,
      text width=1.2cm, align=center,
      font=\sffamily\fontsize{5.5}{7}\selectfont]
    (l3) at (0.85, \ly)
    {\textbf{L3} Agent\\{\fontsize{4.5}{5.5}\selectfont agents, cmds,}\\[-1pt]{\fontsize{4.5}{5.5}\selectfont skills}};

\node[rectangle, rounded corners=2.5pt,
      draw=black!30,
      fill={rgb,255:red,127;green,179;blue,216},
      minimum width=1.45cm, minimum height=1.1cm,
      text width=1.2cm, align=center,
      font=\sffamily\fontsize{5.5}{7}\selectfont]
    (l4) at (2.5, \ly)
    {\textbf{L4} Orchestration\\{\fontsize{4.5}{5.5}\selectfont flows,}\\[-1pt]{\fontsize{4.5}{5.5}\selectfont session-logs}};

\path (0,\ly) coordinate (arrow-y);
\draw[arrthin] (l1.east |- arrow-y) -- (l2.west |- arrow-y);
\draw[arrthin] (l2.east |- arrow-y) -- (l3.west |- arrow-y);
\draw[arrthin] (l3.east |- arrow-y) -- (l4.west |- arrow-y);

\draw[arr] (catout.south) -- ++(0,-0.15);

\pgfmathsetmacro{\bracey}{\ly - 0.55}
\draw[decorate, decoration={brace, mirror, amplitude=3pt},
      black!35, line width=0.4pt]
    (-2.45-0.725, \bracey) -- (2.5+0.725, \bracey)
    node[midway, below=3pt,
         font=\sffamily\itshape\fontsize{5}{6.5}\selectfont,
         text=black!45]
    {Cumulative: max level with $\geq$1 artifact};

\draw[arr] (search.south) -- ++(0,-0.15);

\end{tikzpicture}
\caption{The RAMP classification pipeline.}

\label{fig:ramp-pipeline}
\Description{A vertical three-stage flowchart. Stage~1, artifact discovery, shows a hierarchical search box with three numbered steps---tool patterns for 12 tools, cross-tool patterns in tool directories and the repository root, and non-standard markdown---yielding about 4,300 artifacts before quality filtering. Stage~2, semantic classification, shows nine category templates embedded as 768-dimensional vectors feeding three signal boxes: S1 maps a tool pattern to a category deterministically, S2 embeds path tokens and compares them to the templates by cosine similarity, and S3 does the same with the file content. The three signals merge at a diamond carrying the priority rule S1 over S2-and-S3-agree over S3 over S2, producing one of nine categories per artifact. Stage~3, level assignment, shows four boxes in increasing shade from L1 Unconfigured (no artifacts) through L2 Grounded (rules, config, architecture, style) and L3 Agent (agents, commands, skills) to L4 Orchestration (flows, session logs), under a brace reading that the level is the highest one with at least one artifact.}
\end{figure}

We assign each repository a maturity level through three stages---artifact discovery, semantic classification, and level assignment---summarized in Figure~\ref{fig:ramp-pipeline}.

\smallskip\noindent\textbf{Stage 1: Artifact discovery.}
We search each repository hierarchically: files discovered in earlier steps are not considered later.

\textit{Step~1: Tool-specific standard artifacts.}
We defined 43 file patterns across the 12 AI coding tools with documented conventions for committing configuration to version control (the most widely used such tools at the time of collection): Claude Code (\texttt{.claude/}), Cursor (\texttt{.cursor/}), GitHub Copilot (\texttt{.github/}), Windsurf (\texttt{.windsurf/}), Aider, AMP, Gemini CLI, Google IDX, JetBrains AI, Kiro, OpenAI Codex, and OpenHands.\footnote{The full pattern set is listed in the replication package.}
Patterns match via exact path, glob, or regular expression, and each matched file is attributed to its tool and that tool's artifact category.

\textit{Steps~2a--2b: Cross-tool artifacts.}
We then search for cross-tool patterns (\texttt{AGENTS.md}, \texttt{*mcp*.json}) first inside tool configuration folders, then at the repository root.
Root-level files receive \texttt{tool=shared}, since a root-level \texttt{AGENTS.md} follows cross-tool interoperability standards and belongs to no single tool.

\textit{Step~3: Non-standard markdown.}
Remaining markdown files are swept for discovery statistics only; classification evidence comes solely from validated artifacts.

Filename patterns cannot capture the full semantic range of AI configuration files---a \texttt{.cursorrules} and a \texttt{CLAUDE.md} file may encode the same behavioral rules under different names---so Stage~2 classifies each artifact by content.

\smallskip\noindent\textbf{Stage 2: Semantic classification.}\label{sec:classification}
Each discovered artifact is classified into one of nine semantic categories (Table~\ref{tab:levels}) using a priority-ordered ensemble of three independent signals.

\textit{Embedding model and category templates.}
We embed text with \texttt{nomic-ai/nomic-embed-text-v1.5}, a sentence transformer whose 768-dimensional vectors and 8,192-token window accommodate long configuration files (some exceed 200 lines) without truncation.
Each category is defined by a 40--120-word template capturing its archetype, with negative clauses that maximize separation in embedding space; embedding the templates yields nine reference vectors for comparison (Table~\ref{tab:templates}).

\begin{table*}[t]
\caption{The nine RAMP category templates, their maturity-level assignments, and the general-documentation fallback. Each template is embedded once; artifacts are classified by cosine similarity against these reference vectors (Stage~2). The general-documentation template absorbs non-AI documentation and contributes no level.}
\label{tab:templates}
\footnotesize
\setlength{\tabcolsep}{4pt}
\renewcommand{\arraystretch}{1.12}
\begin{tabular}{p{2.2cm}cp{13.6cm}}
\toprule
\textbf{Category} & \textbf{Level} & \textbf{Template text (embedded verbatim)} \\
\midrule
rules & L2 & A policy document of imperative directives that govern how an AI assistant must behave in a codebase. Uses mandatory language like NEVER, ALWAYS, MUST, and DO NOT to enforce constraints and conventions. Does not contain code examples, workflow orchestration tables, or step-by-step tutorials — only behavioral rules and project-level instructions. \\
\midrule
configuration & L2 & A machine-readable JSON, YAML, or TOML file with hierarchical key-value pairs, boolean flags, and nested settings objects. Defines tool servers, environment variables, permission scopes, file patterns, and feature toggles. Contains no prose paragraphs or natural language instructions — purely structured data for tool or environment configuration. \\
\midrule
architecture & L2 & A system design document describing software architecture with component diagrams, data flows, and deployment topology. Uses Mermaid, PlantUML, or ASCII diagrams with ADR-style decision records and C4 model levels. Covers infrastructure, service boundaries, scaling strategies, and technology stack rationale — not coding standards or runtime configuration. \\
\midrule
code-style & L2 & A coding standards document with before-and-after code comparisons showing incorrect vs correct patterns. Contains inline code examples, linting rules, type safety requirements, naming conventions with specific casing, and coverage metrics. Focuses on how source code should be written at the syntax level — unlike behavioral rules which govern AI assistant conduct. \\
\midrule
agents & L3 & A persona definition file that establishes an AI agent's identity, role, and behavioral boundaries. Contains YAML frontmatter with structured fields like name, type, model, tools, and capabilities. Defines delegation boundaries, domain expertise scope, and interaction protocols for a single autonomous agent. \\
\midrule
commands & L3 & A short, self-contained prompt template that defines exactly one executable action a user can invoke. Typically under 25 lines with a slash-command trigger, parameterized \$ARGUMENTS, and a single output. Not a multi-step orchestration or policy document — just one atomic, reusable operation like commit, review, or format. \\
\midrule
skills & L3 & A long-form how-to guide (typically 200--600 lines) that teaches a specific technique or capability in depth. Includes trigger conditions, detailed step-by-step methodology, MCP tool usage, edge case handling, and validation criteria. Functions as reusable domain expertise that can be composed and extended, unlike short commands or behavioral rules. \\
\midrule
flows & L4 & An executable orchestration plan authored to be consumed by an AI runner — names specific phases with identifiers, assigns named agents or workers to each phase, declares explicit dependency edges between stages, and lists per-phase acceptance/exit criteria. The file directly drives execution: a runner could parse it and dispatch work. NOT a design document explaining how orchestration works, NOT a tutorial teaching about workflows, NOT a roadmap or release schedule, NOT a code recipe describing how an event sequence behaves, NOT a doc page about multi-agent systems — those describe orchestration; this IS an orchestration. \\
\midrule
session-logs & L4 & An actual log entry produced by a specific AI agent run — captures concrete artifacts of that run: a run/session/task identifier, real timestamps of state transitions that occurred, the names of files that were actually modified, real commit SHAs that were made, the agent's actor identity, and the outcome of acceptance criteria. The file is the OUTPUT of an executed agent run — a forensic record that exists because the run happened. NOT a documentation page describing how session logs work, NOT a tutorial about agent memory or logging, NOT an observability guide, NOT a state-machine reference manual, NOT a tracing-system explanation, NOT a retrospective design discussion, NOT an event-sourcing recipe — those describe logs; this IS a log. \\
\midrule
general-documentation & --- & User-facing software project documentation written for end users, contributors, or operators of a non-AI software system. Covers installation, API reference, usage tutorials, performance characteristics, troubleshooting, FAQs, design rationale, deployment guides, and operational concerns. Describes how the project itself works — does not configure, instruct, or orchestrate any AI assistant or agent. Not authored by an AI tool, not consumed by an AI tool: ordinary technical writing for humans about a software product. \\
\bottomrule
\end{tabular}
\end{table*}

\textit{Three signals, combined by priority.}
\emph{(1)~Tool-pattern matching} (highest priority): for artifacts found in Step~1, we take the category from the tool's configuration, normalizing ``instructions'' files (e.g., \texttt{CLAUDE.md}, \texttt{copilot-\allowbreak instructions.md}) to \texttt{rules}.
\emph{(2)~Path semantics}: the tokenized path (e.g., \texttt{.claude/commands/sparc.md} $\rightarrow$ ``claude commands sparc'') is embedded and compared by cosine similarity to the nine templates.
\emph{(3)~Content semantics}: the full text is embedded and compared likewise, with no minimum similarity threshold imposed (Section~\ref{sec:validation} reports what happens when one is).
When a tool category is present we use it; otherwise the category on which path and content agree; on disagreement, the content signal (richer than a tokenized path); and the path signal only when content is unavailable.
Content alone matches the full pipeline for 42.4\% of repositories; the rest benefit from the additional signals.

\textit{Worked example.}
In a repository with \texttt{CLAUDE.md}, \texttt{.cursorrules}, and \texttt{.claude/agents/security-reviewer.md}, the tool-pattern matches classify directly (two \texttt{rules}, one \texttt{agents}); a fourth file, \texttt{docs/migration-guide.md}, has no tool signal and its content---step-by-step procedures with validation criteria---embeds closest to \texttt{skills}, resolving the repository to \textbf{Level~3}.

\smallskip\noindent\textbf{Stage 3: Level assignment.}
The nine categories map to four cumulative maturity levels (Table~\ref{tab:levels}), and a repository's overall level is the highest level at which it has at least one classified artifact.
Higher-level repositories are expected to contain lower-level artifacts---an assumption we test empirically below rather than impose.
A Level~4 repository lacking Level~2 artifacts triggers a coherence warning and a reduced confidence score but is still assigned Level~4, separating classification (what is present) from validation.

\subsection{Validation}
\label{sec:validation}

We validate the RAMP classification by characterizing the resulting landscape, testing whether the four levels form a genuine cumulative scale and whether the specific category-to-level mapping is justified, validating the labels against independent human annotation, and assessing robustness to tool choice, repository characteristics, and classification parameters.

\subsubsection{Adoption Landscape}

Across the 441 repositories, adoption is dominated by Level~1: 294 repositories (66.7\%) show no evidence of structured practice, 109 (24.7\%) are at Level~2 (grounded prompting), 38 (8.6\%) at Level~3 (agent-augmented), and none at Level~4.
Artifact volume scales sharply with maturity---Level~1 repositories average 0.5 validated artifacts, Level~2 repositories 2.7 (median 1), and Level~3 repositories 15.8 (median 8.5)---so higher maturity reflects materially richer configuration.

\begin{figure}[t]
\centering
\includegraphics[width=\columnwidth, clip=true, trim=10 25 15 10]{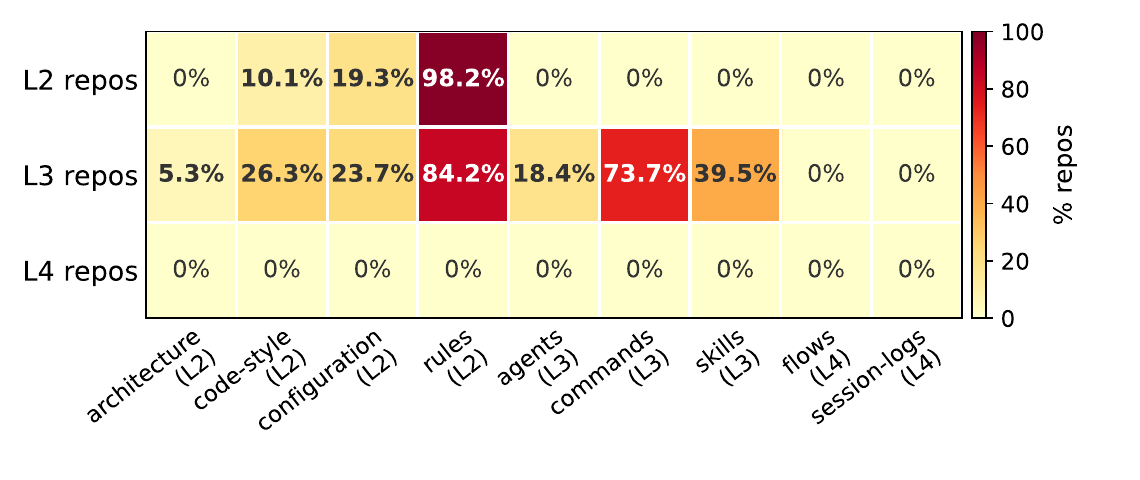}
\caption{Category prevalence by maturity level. Rules files are near-universal at both observed levels; Level~3 categories appear only at Level~3; Level~4 categories are absent.}
\label{fig:heatmap}
\Description{A heatmap with three rows---Level~2, Level~3, and Level~4 repositories---and nine category columns ordered architecture, code-style, configuration, and rules (Level~2), then agents, commands, and skills (Level~3), then flows and session-logs (Level~4). Each cell gives the share of repositories at that level containing the category, shaded from pale yellow at zero to dark red at 100 percent. The Level~2 row reads 0, 10.1, 19.3, and 98.2 percent across the Level~2 categories and zero everywhere else. The Level~3 row reads 5.3, 26.3, 23.7, and 84.2 percent across the Level~2 categories, 18.4, 73.7, and 39.5 percent across the Level~3 categories, and zero for the Level~4 categories. The Level~4 row is zero throughout.}
\end{figure}

The number of distinct categories a repository uses rises with its maturity level (Spearman $\rho = 0.843$; $\rho = 0.727$ on the AI-tools subset, the more conservative estimate since Level~1 repositories have zero breadth by construction).
Rules files are near-universal: 98.2\% of Level~2 and 84.2\% of Level~3 repositories have one.
Level~3 categories appear only at Level~3, and Level~4 categories not at all (Figure~\ref{fig:heatmap}), confirming the layered structure.
Repositories combine categories in only a few ways: the 210 scored repositories account for just 25 distinct combinations, and those combinations form tight, well-separated groups (silhouette 0.822 against a permutation null of 0.731, $p < 0.001$).
The most common combination is a single rules file.
Each group adds categories to the one below it, so the groups line up with the maturity scale; none is organized around a specialty, such as architecture documents committed without rules.

\subsubsection{Cumulative Structure}

We test whether the four maturity levels form a cumulative hierarchy using Guttman scalogram analysis~\cite{guttman1971measurement}, coding each repository as a binary vector indicating whether it has primary artifacts at Level~2, Level~3, and Level~4.
The coefficient of reproducibility (CR) measures the fraction of binary responses conforming to the cumulative ordering; the coefficient of scalability (CS) adjusts for item prevalence; standard thresholds are CR $\geq$ 0.90 and CS $\geq$ 0.60.
Both are exceeded (CR = 0.997, CS = 0.983), confirming that the levels conform to a cumulative scale.
Direct inspection reinforces this: 89.5\% of Level~3 repositories have primary Level~2 evidence, and 100\% when secondary evidence is included.
On the 217-repository AI-tools subset the scale still passes comfortably (CR = 0.994, CS = 0.937); the violation rate is 0.9\% on the full frame and 1.8\% on the subset.
Only four repositories violate strict cumulative ordering, all with Level~3 evidence but no Level~2 grounding.
With no Level~4 evidence observed, the Guttman certificate covers only the Level~2 to Level~3 transitions; the mapping-permutation test and prevalence ordering below establish validity for the full hierarchy.
As an honest calibration, the mean Loevinger $H = 0.299$: the hierarchy is strong at the aggregate level but only weakly expressed as item-pair implications in this sparse population.

\subsubsection{Mapping Validation}

To test whether any grouping of the nine categories into three levels would produce a valid scale, we shuffled them into random level assignments 10,000 times and recomputed the Guttman metrics.
The observed CR falls at the 99.3rd percentile of the null distribution ($p = 0.0074$); the observed CS at the 99.5th percentile ($p = 0.0051$).
The Kendall rank correlation between category prevalence and assigned level ($\tau = 0.663$, $p = 0.0235$) confirms that more prevalent categories are assigned to lower levels, as expected for a cumulative scale.
All nine leave-one-out variants pass both thresholds, confirming that the scale does not depend on any single category.

\subsubsection{Human Validation}
\label{sec:human-validation}

\begin{table}[t]
\centering
\caption{Agreement between the RAMP classifier and human judgment on a held-out, level-stratified sample (three annotators, 195 labeled artifacts across 35 repositories).}
\label{tab:human-validation}
\small
\begin{tabular}{lr}
\toprule
Inter-annotator agreement (Krippendorff's $\alpha$) & 0.572 \\
\midrule
\multicolumn{2}{l}{\textit{File level (175 reference items)}} \\
\quad Accuracy vs.\ human reference (Wilson 95\% CI) & 81.7\% [75.3, 86.7] \\
\quad Accuracy on majority-vote references ($n = 47$) & 93.6\% \\
\quad Cohen's $\kappa$ vs.\ human reference & 0.743 \\
\midrule
\multicolumn{2}{l}{\textit{Repository level (35 repositories)}} \\
\quad Exact maturity-level match & 34/35 (97.1\%) \\
\quad Quadratic-weighted $\kappa$ & 0.829 \\
\bottomrule
\end{tabular}
\end{table}

The structural tests above certify the scale; to validate the labels themselves, three annotators independently labeled 195 artifacts from 35 repositories of the Study~2 application corpus (held out from the development sample), sampled in level-balanced blocks so that rare higher-level strata are represented.
Annotators were blinded to the machine labels and made a forced choice among the nine categories plus general-documentation and not-an-artifact, under a partial-overlap design (128 items labeled by one rater, 59 by two, 8 by all three; 270 labels total).
Reference labels are the majority vote where available and the single rater's label otherwise; 20 items without a majority are excluded, leaving 175 reference items.

Inter-annotator agreement is moderate (Krippendorff's $\alpha = 0.572$; pairwise Cohen's $\kappa$ 0.50--0.81): the category boundaries are genuinely hard to separate, and human agreement bounds how sharply any classifier's accuracy on this task can be read.
Against the human reference the classifier agrees on 81.7\% of items (Wilson 95\% CI 75.3--86.7; Cohen's $\kappa = 0.743$), inside the human--human agreement range, rising to 93.6\% on the majority-backed subset where the human signal is strongest; an independent LLM-based labeler (Claude Haiku) evaluated on the same items is statistically indistinguishable (81.1\%; exact McNemar $p = 1.0$), suggesting that much of the residual disagreement is irreducible label ambiguity rather than classifier idiosyncrasy.

Most importantly for both studies, propagating the evaluated files through the production level-assignment rule reproduces the human-derived repository maturity level for 34 of 35 repositories (97.1\%; quadratic-weighted $\kappa = 0.829$; Table~\ref{tab:human-validation}).
Disagreements concentrate in the rare procedural categories (skills, flows; human support $\leq 5$ items each), where the classifier tends to fall back to rules or discard the artifact and where annotators disagree most; because the sample is level-stratified, the reported rates are stratum-weighted rather than population rates.

\subsubsection{Robustness}

\textit{Tool-specific bias.}
Maturity distributions differ across the three major tools (chi-squared = 111.3, $p < 0.0001$) at medium effect size (Cramer's V = 0.483): GitHub Copilot repositories concentrate at Level~1, Cursor repositories at Levels~2--3.
All three tools exhibit the L2$\rightarrow$L3 progression.
The divergence partly reflects measurement asymmetry across tool ecosystems: the validated evidence standard is most visible for tools whose conventions mandate committed configuration files, a first-order caveat for cross-tool comparisons.

\textit{Repository characteristics.}
Eight repository features show weak-to-moderate raw correlations with maturity level.
After controlling for artifact count, size measures collapse (log lines: partial $\rho = 0.012$), but commit velocity retains a partial correlation of $\rho = 0.269$ ($p \approx 8 \times 10^{-5}$).
Maturity is not a proxy for project size; repository activity---or the engineering discipline it proxies---retains an independent association and should be controlled in cross-sectional comparisons.

\textit{Classification thresholds.}
Imposing a minimum cosine similarity on the content signal, swept from 0.00 to 0.40 across 21 values, leaves the Guttman CR at $\geq 0.90$ throughout.
The sweep is uninformative, however: no threshold in this range excludes a single artifact or moves a single repository.

\section{Study 1: Adoption Dynamics (RQ1)}
\label{sec:study1}

\begin{figure}[t]
\centering
\includegraphics[width=0.8\columnwidth]{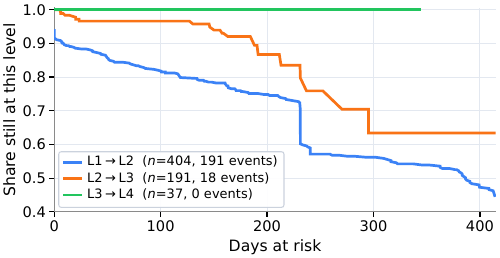}
\caption{Kaplan-Meier curves for time to the next level; each clock starts on entry to that level.}
\label{fig:km}
\Description{A line chart with three Kaplan-Meier survival curves. The horizontal axis is days at risk, from 0 to about 415; the vertical axis is the share of repositories still at that level, from 0.4 to 1.0. The blue L1-to-L2 curve, 404 at risk with 191 events, starts at 0.94 and declines steadily, drops sharply from 0.72 to 0.60 at about 230 days, and ends near 0.45. The orange L2-to-L3 curve, 191 at risk with 18 events, holds near 0.97 until about 130 days, steps down to 0.63 by 300 days, and stays flat thereafter. The green L3-to-L4 curve, 37 at risk with no events, is flat at 1.0 across its whole range, ending at about 345 days.}
\end{figure}

Having validated RAMP as a cumulative maturity scale, Study~1 asks how repositories move through the maturity levels over time.
\label{sec:rq1_results}
We date every validated artifact by its first commit and mark a level transition when an artifact of a new level first appears, over git events from Jan~2025 to Feb~2026 (about 14 months; Section~\ref{sec:threats} discusses the short window).
Transition timing comes from Kaplan-Meier curves (Figure~\ref{fig:km}) estimated on the full 441-repository frame, with repositories that never commit a validated artifact entering as right-censored observations, so Level~1 durations are lower bounds; the transition statistics below condition on the 196 adopters.
See replication package for at-risk counts, censoring rules, and per-curve clocks.

Adoption is single-shot, forward-only, and front-loaded, and the hardest transition is the first one.
Almost every adopter enters at Level~2 (95.9\%; the remainder go directly to Level~3), after a median of at least 441 observed days with no validated artifact (timed from first observed activity; 633 days among the 87\% with measurable pre-adoption history).
After that first artifact most repositories stop: 83.1\% introduce artifacts at exactly one level, 14.9\% progress from Level~2 to Level~3, and 2.1\% follow inverted paths.
Reversals are absent (0\%) and abandonment negligible (0.5\%), but 12.8\% of non-transitioning repositories sit in a plateau, concentrated at Level~2.
The Level~2 to Level~3 transition, when it happens at all, is comparatively fast: a median of 154 days ($n = 18$) against the 633-day latency to a first artifact.
Reaching the first structured artifact takes years; the next transition, when it happens, takes months.

At the artifact level, 73.8\% of configurations follow a set-and-forget lifecycle: committed once and never modified.
Maintenance effort appears to scale with maturity, but the association disappears once artifact count is controlled (Kruskal-Wallis $H = 60.61$, $p < 0.0001$; partial $\rho = -0.072$, $p = 0.32$).
The one metric that does scale is author diversity: Level~3 repositories involve a mean of 3.4 unique contributors to AI artifacts, compared to 1.0--1.5 at Levels~1--2 ($\rho = 0.402$).
Agent-level adoption is a team activity.

The timing also supports a developmental ordering: transitions that introduce the lower level first outnumber the reverse by roughly 4.5:1 ($\rho = 0.866$; exact one-sided permutation $p = 0.029$).
About a third of multi-level repositories introduce both levels in the same commit or day, however, consistent with teams configuring AI tools in cumulative bursts rather than through sequential stages.

AI artifacts are configured once and left in place.
Higher maturity reflects broader initial investment, with no corresponding increase in ongoing stewardship.

\section{Study 2: Agent Adoption Effects by Maturity}
\label{sec:study2}

\subsection{Data, Treatment, and Sample}
\label{sec:study2_data}

Agarwal et al.~\cite{agarwal2026} estimated the causal effects of autonomous coding agent adoption across open-source repositories using staggered difference-in-differences with propensity-score matched controls.
The unit is the repository-month; treatment is the first agent-attributed pull request, identified through a cascade of branch prefixes (e.g., \texttt{cursor/}, \texttt{claude/}), author logins, commit metadata, and PR description patterns.
Their sample comprises 401 agent-first repositories matched to 606 controls and 117 IDE-first repositories matched to 73 controls, to which we add our maturity stratification.

Of the 518 treated repositories, 9 could not be traced at re-collection (deleted, made private, or renamed), leaving 509 for stratified analysis: 236 at Level~1 (no committed AI configuration) and 273 at Level~2+ (210 L2, 23 L3, 40 L4).
Unlike the corporate development sample of Section~\ref{sec:ramp_data}, the open-source treated sample does contain orchestration-level (L4) evidence.
Table~\ref{tab:l1_vs_l2} compares the two groups on pre-treatment characteristics.
Level~2+ repositories match Level~1 in age but are far more starred and more active in forks and pull requests, consistent with established projects investing more in AI configuration.
Difference-in-differences removes time-invariant confounders; residual confounding from correlated engineering practices is discussed in Section~\ref{sec:threats}.

\begin{table}[t]
\caption{Pre-adoption characteristics of repositories by AI practice maturity (Level 1 vs Level 2+).}
\label{tab:l1_vs_l2}
\small
\setlength{\tabcolsep}{3pt}
\begin{tabular}{p{1.5cm}cccccccc}
\toprule
& \multicolumn{2}{c}{Mean}
& \multicolumn{2}{c}{Min}
& \multicolumn{2}{c}{Median}
& \multicolumn{2}{c}{Max} \\
\cmidrule(lr){2-3} \cmidrule(lr){4-5} \cmidrule(lr){6-7} \cmidrule(lr){8-9}
Characteristic
& L1 & L2+
& L1 & L2+
& L1 & L2+
& L1 & L2+ \\
\midrule
Age (days)$^{*}$         & 1,042 & 1,055 & 0 & 0 & 553.5 & 639 & 5,250 & 5,221 \\
Stars$^{\dagger\dagger}$              & 831 & 4,907 & 10 & 10 & 55 & 174 & 26,818 & 177,379 \\
Forks$^{\dagger\dagger}$              & 171.7 & 773.3 & 0 & 0 & 10 & 38 & 6,026 & 45,908 \\
PRs$^{\dagger}$            & 639.7 & 1,358 & 10 & 10 & 87 & 389 & 31,836 & 35,389 \\
Agentic PRs$^{\dagger}$     & 109.4 & 123.5 & 10 & 10 & 23 & 39 & 13,462 & 4,023 \\
\bottomrule
\multicolumn{9}{l}{\footnotesize $^{*}$ At agent adoption. \quad $^{\dagger}$ As of the end-November 2025 snapshot.} \\
\multicolumn{9}{l}{\footnotesize $^{\dagger\dagger}$ As reported in the original dataset.} \\
\end{tabular}
\end{table}

\subsection{Estimation, Outcomes, and Moderation}
\label{sec:moderator}

We adopt Agarwal et al.'s~\cite{agarwal2026} estimator without modification: staggered difference-in-differences via the Borusyak et al.~\cite{borusyak2021} imputation estimator, with standard errors clustered by repository.
Velocity is measured with monthly commits and lines of code added; software quality with four SonarQube~\cite{sonarqube} metrics (cognitive complexity, static-analysis warnings, duplicated line density, and comment line density).
This outcome set is Agarwal et al.'s, adopted unchanged so our stratified estimates remain directly comparable to their pooled ones; the quality metrics are standard static maintainability indicators and match those used in prior longitudinal work on AI-assisted coding~\cite{he2026cursor}.
All outcomes are log-transformed, so coefficients approximate percentage changes.
Increases in complexity, warnings, and duplication indicate quality degradation; higher comment density indicates improved documentation.

We treat maturity as a moderator: we estimate agent-adoption effects separately within each maturity stratum and compare them.
Any difference reflects heterogeneity in the treatment effect of adoption, not a causal effect of maturity itself.

Our primary analysis collapses maturity into two groups, Level~1 versus Level~2+, and estimates within the agent-first stratum, which holds prior AI exposure constant.
Agent-first and IDE-first repositories differ at baseline and respond differently to adoption~\cite{agarwal2026}, so pooling them would confound exposure with maturity, and the IDE-first Level~1 stratum holds only 22 repositories---too few to identify the contrast on its own (Appendix~\ref{app:crosstab}).
IDE-first estimates are reported in the replication package.

We assign maturity labels using the RAMP classifier applied to each repository as of November 2025, the terminal date of Agarwal et al.'s panel.
Study~1 shows that most AI artifacts follow a set-and-forget lifecycle (Section~\ref{sec:rq1_results}), so an end-of-window snapshot is an informative proxy for stable practice.
We partition treated repositories by maturity after Agarwal et al.'s matching step, so Level~1 and Level~2+ units share the same control pool.
Two limitations follow from this design: some Level~2+ repositories may have committed artifacts only after agent adoption (reverse causality), and the shared control pool does not guarantee covariate balance within strata.
Both are discussed in Section~\ref{sec:threats}.

We assess robustness four ways (reported in Section~\ref{sec:robustness}): event-study pre-trends to check parallel trends before adoption; the full four-level maturity gradient; a combined Level~1--2 grouping to test the binary boundary; and the pooled agent-first/IDE-first sample as a dilution check.
We also examine whether the gradient reflects differential treatment intensity (agent PR volume) rather than differential response, addressed qualitatively in Section~\ref{sec:threats}.

\begin{figure}[t]
\centering
\includegraphics[width=0.8\columnwidth,clip=true,trim=0 10 0 15]{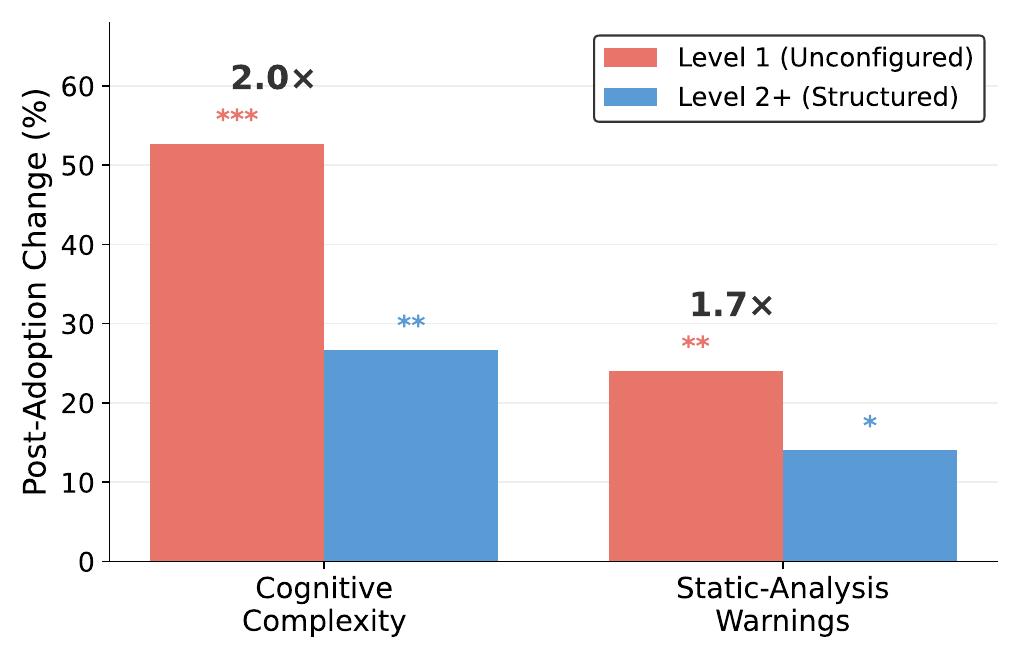}
\caption{Quality degradation after agent adoption by RAMP maturity in the agent-first stratum (higher = worse). Level~1 repositories show 1.7--2.0$\times$ larger increases than Level~2+. $^{*}p<0.05$; $^{**}p<0.01$; $^{***}p<0.001$.}
\label{fig:gradient}
\Description{A grouped bar chart with two metric groups, cognitive complexity and static-analysis warnings, and two bars per group: red for Level~1 (unconfigured) and blue for Level~2+ (structured). The cognitive complexity bars reach about 53 and 27 percent and are annotated 2.0 times; the static-analysis warning bars reach about 24 and 14 percent and are annotated 1.7 times. Significance stars sit above each bar: three for Level~1 complexity, two for Level~2+ complexity, two for Level~1 warnings, and one for Level~2+ warnings.}
\end{figure}

\subsection{RQ2: Associations With Outcomes}

We now test whether RAMP maturity is associated with different outcomes when repositories adopt coding agents.

\subsubsection{Quality Effects}

Practice maturity strongly moderates the quality impact of agent adoption (Table~\ref{tab:main_results}; Figure~\ref{fig:gradient}).

\begin{table}[t]
\caption{Average post-adoption effects of coding agent adoption in the agent-first stratum (primary specification), stratified by AI practice maturity. All outcomes log-transformed.}
\label{tab:main_results}
\small
\setlength{\tabcolsep}{1pt}
\begin{tabular}{p{2cm}ccccc}
\toprule
& \multicolumn{2}{c}{Level 1 (Unconfigured)} & \multicolumn{2}{c}{Level 2+ (Structured)} & \\
\cmidrule(lr){2-3} \cmidrule(lr){4-5}
Outcome & $\beta$ (SE) & \% $\Delta$ & $\beta$ (SE) & \% $\Delta$ & Ratio \\
\midrule
\multicolumn{6}{l}{\textit{Quality}} \\
\quad Cognitive\newline complexity & $0.423^{***}$ (0.118) & +52.70\% & $0.236^{**}$ (0.077) & +26.68\% & 2.0$\times$ \\
\quad Static-analysis warnings & $0.216^{**}$ (0.077) & +24.08\% & $0.131^{*}$ (0.066) & +14.04\% & 1.7$\times$ \\
\quad Duplicated\newline line density & $0.144^{*}$ (0.073) & +15.44\% & $0.133^{**}$ (0.049) & +14.23\% & 1.1$\times$ \\
\quad Comment\newline line density & $0.068$ (0.044) & +7.07\% & $0.019$ (0.029) & +1.91\% & 3.7$\times$ \\
\midrule
\multicolumn{6}{l}{\textit{Velocity}} \\
\quad Commits & $0.319^{***}$ (0.074) & +37.56\% & $0.243^{***}$ (0.067) & +27.52\% & 1.4$\times$ \\
\quad Lines added & $0.393^{*}$ (0.191) & +48.07\% & $0.523^{***}$ (0.133) & +68.72\% & 0.7$\times$ \\
\bottomrule
\multicolumn{6}{l}{\footnotesize $^{*}p<0.05$;\quad $^{**}p<0.01$;\quad $^{***}p<0.001$.} \\
\multicolumn{6}{l}{\footnotesize Ratio is the L1 percentage change divided by the L2+ percentage change.} \\
\end{tabular}
\end{table}

Level~1 repositories show larger quality degradation than Level~2+ on every metric: cognitive complexity roughly doubles the Level~2+ increase (+52.7\% versus +26.7\%) and static-analysis warnings are 1.7$\times$ larger (+24.1\% versus +14.0\%).
Duplicated-line density differs little between strata within the agent-first sample (+15.4\% versus +14.2\%), though in the pooled sample the Level~2+ increase is near zero (+5.9\%, n.s.) against +16.0\% at Level~1.
Even with structured practices, a significant 26.7\% complexity increase persists.

\begin{figure*}[t]
\centering
\includegraphics[width=\textwidth,clip=true,trim=0 10 0 0]{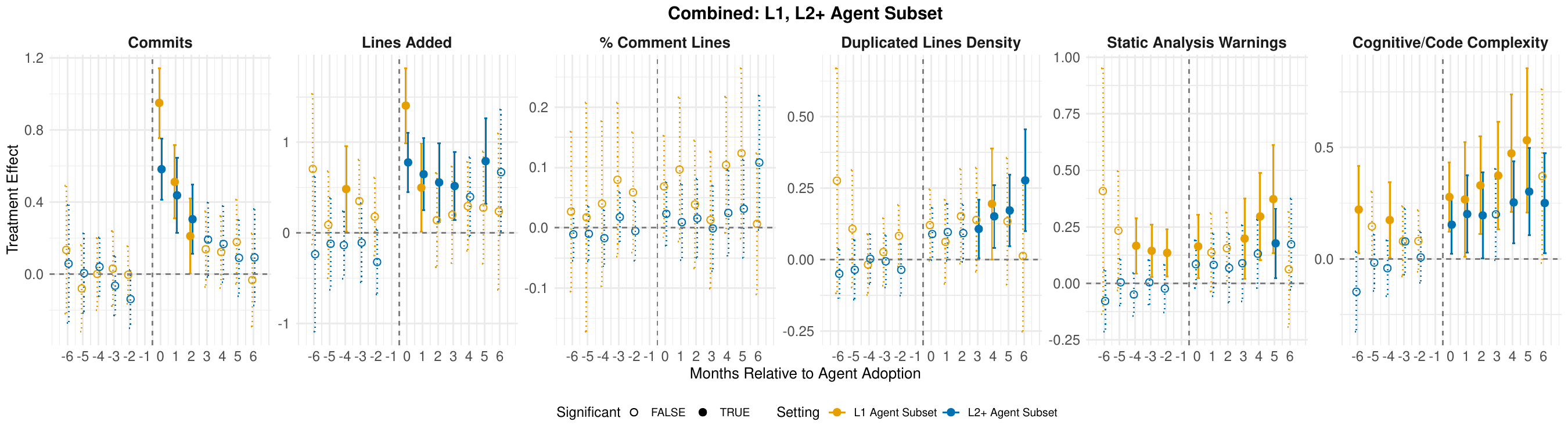}
\caption{Dynamic treatment effects by months relative to the first agent-attributed pull request, split by RAMP maturity.
Quality metrics show the Level~1 versus Level~2+ gap widening after adoption; velocity paths remain aligned across strata.}
\label{fig:event_study}
\Description{A row of six event-study panels sharing a horizontal axis of months relative to agent adoption, from minus six to plus six, with a dashed vertical line at adoption and a dashed horizontal line at zero effect. The panels are, left to right, commits, lines added, percent comment lines, duplicated line density, static-analysis warnings, and cognitive complexity. Each panel plots two series of point estimates with 95 percent confidence whiskers, orange for the Level~1 agent subset and blue for the Level~2+ agent subset, with filled markers for significant estimates and hollow markers otherwise. Pre-adoption estimates sit near zero in both series across panels, except in the static-analysis warnings panel, where the orange Level~1 series is already elevated. After adoption the orange series rises above the blue in the warnings and complexity panels and the gap widens through month six, while the commits and lines-added panels show the two series overlapping.}
\end{figure*}

For context, He et al.~\cite{he2026cursor} reported 30\% warning and 41\% complexity increases for Cursor adopters overall.
Level~2+ repositories fall below both baselines; Level~1 exceeds the complexity baseline but stays below it for warnings.
The aggregate that prior work reports averages over these divergent trajectories.

One caveat applies to this comparison: He et al.\ studied a single IDE tool (Cursor) in an earlier period, whereas our sample spans many autonomous agents in a later window when models and tooling had evolved, and Level~2+ repositories may disproportionately use more capable frontier models.
Whether the lower Level~2+ effects partly reflect model-capability gains rather than practice is an alternative our design cannot rule out---a first-order interpretive limitation of Study~2 (Section~\ref{sec:threats}).

\subsubsection{Velocity Effects}

Commit frequency increases significantly in both strata: +37.6\% at Level~1 and +27.5\% at Level~2+.
Lines of code added run the other way: Level~2+ repositories show a +68.7\% increase ($p < 0.001$) against +48.1\% at Level~1 ($p < 0.05$).
Neither velocity metric shows the systematic Level~1 penalty that characterizes complexity and warnings; if anything, structured repositories convert agent activity into more code.

\subsubsection{Temporal Dynamics}

Figure~\ref{fig:event_study} plots dynamic treatment effects for all outcomes, stratified by RAMP maturity.

For quality outcomes, Level~1 series generally rise through the post-adoption window with little sign of returning toward pre-adoption levels, whereas Level~2+ trajectories show an initial bump followed by moderation; the gap widens rather than closes over event time.
Velocity panels remain aligned across strata, matching the static velocity effects in Table~\ref{tab:main_results}.
Pre-treatment coefficients for cognitive complexity are small relative to post-adoption effects in both strata; for static-analysis warnings the Level~1 stratum shows elevated pre-treatment coefficients, so the warnings contrast is more weakly identified than complexity (Section~\ref{sec:threats}).

\subsubsection{Robustness}
\label{sec:robustness}

\textbf{Gradient decomposition.} Estimating effects separately for each of the four maturity levels (pooled sample) confirms a monotonic decline: cognitive complexity rises 54.4\% at Level~1 and 39.8\% at Level~2 (both $p < 0.001$) and is indistinguishable from zero at Levels~3--4, static-analysis warnings follow the same ordering (+23.0\% at Level~1, +16.7\% at Level~2, non-significant above), and duplicated-line density is significant only at Level~1 among Levels~1--3.
Level~1 shows the largest degradation on every quality metric, though the Level~3 and Level~4 strata are small (23 and 40 repositories).

\textbf{Level~4 dose-response.} Level~4 is a partial exception to the gradient: duplicated-line density rises by 31.4\% ($p < 0.01$), the only significant Level~4 quality effect, consistent with a dose-response pattern in which orchestration-level repositories are the heaviest agent users and accumulate duplicated code even as complexity remains statistically flat.

\textbf{Boundary sensitivity.} In the pooled sample, the combined Level~1--2 group shows intermediate effects: cognitive complexity increases by 46.6\% and static-analysis warnings by 20.9\% (both $p < 0.001$), between the Level~1 and Level~2+ estimates.
The ordering Level~1 $>$ Level~1--2 $>$ Level~2+ holds for complexity and warnings.

\textbf{Pooled-sample dilution.} Pooling agent-first and IDE-first repositories compresses the complexity gradient (+54.4\% versus +35.9\%, 1.5$\times$) and nearly erases the warnings gradient (+23.0\% versus +20.0\%), as the heavily configured IDE-first population enters almost entirely on the Level~2+ side, reinforcing the agent-first stratum as the primary specification.

Across all checks the asymmetry holds: agents accelerate both strata, but the steepest quality costs concentrate in repositories without structured practices.

\section{Discussion}
\label{sec:discussion}

\subsection{Interpretation}

Read together, the two studies link how teams configure AI to what happens when they deploy it. Study~1 shows that AI configuration is a cumulative but largely one-time investment: movement through the levels is fast when it occurs, yet most repositories plateau at their entry level, and 73.8\% of artifacts are committed once and never revised. Study~2 shows that this configuration predicts outcomes: agents accelerate development at every maturity level, but in the agent-first stratum the complexity cost at Level~1 is about twice that at Level~2+.

The gradient is clearest for cognitive complexity and static-analysis warnings (+24.1\% versus +14.0\%); the duplication contrast between strata appears only in the pooled sample. This decomposition changes how the aggregate literature should be read: Agarwal et al.~\cite{agarwal2026} report an average complexity increase of 39\% across all adopters, but within the agent-first stratum that average separates into +52.7\% at Level~1 and +26.7\% at Level~2+, describing neither group.

Three channels could produce the gradient, and they are not mutually exclusive.
First, committed artifacts may act as direct guardrails on agent output: a \texttt{.cursorrules} file specifying ``Always use named constants; never use magic numbers'' or ``Run \texttt{eslint} before suggesting changes are complete'' constrains every agent-generated contribution and lowers static-analysis warnings by construction.
Second, the artifacts may be markers of broader engineering discipline, since teams that commit AI configuration may also review more strictly and test more thoroughly.
Third, structured practices may speed organizational learning by giving a team a shared, updatable reference for working with agent-generated code.
Our design cannot distinguish among them.

The guardrail channel gets weak but specific support from dupli\-cated-line density in the pooled sample: Level~2+ repositories show a much smaller increase (+5.9\%, n.s.) than Level~1 (+16.0\%), consistent with rules and coding-standards files enforcing conventions that prevent repeated code, though the contrast does not persist within the agent-first stratum.
Practitioner accounts point the same way: developers who report the largest gains from agentic workflows credit deliberate practice over the tools: Ronacher~\cite{ronacher2025} attributes his results to keeping code simple enough for agents to work in and tightly limiting what they may change, and Willison~\cite{willison2026} treats agentic engineering as a set of learnable patterns.
Such accounts support our hypothesis but cannot measure the practices they credit.
Whichever channel dominates, the implication is the same: committed structured practices are associated with substantially lower quality degradation.

AI governance extends well beyond the repository: teams also encode practice in wikis, onboarding material, review checklists, and pull-request gates, none of which our pipeline can see. A repository's RAMP level is therefore a lower bound on its organization's maturity, so some repositories we label Level~1 are well governed by means we cannot observe---a misclassification that attenuates the reported gap rather than inflating it (Section~\ref{sec:threats}).

\subsection{Practical Implications}

For practitioners, the gap that matters most is between Level~1 and Level~2. Reaching the first committed artifact takes years, but the next level takes only months (Section~\ref{sec:rq1_results}), and Level~2 is inexpensive: behavioral rules, coding standards, and basic configuration files, often a few pages of committed markdown. Their presence is associated with roughly half the quality degradation of their absence, and higher levels add only smaller reductions.

Concretely, teams should commit at least a rules file and a coding-standards document before or during the first months of agent deployment. Because most AI artifacts are committed once and never modified, that initial configuration will likely govern agent behavior for the life of the project. Tooling that lowers the cost of the first configuration, through sensible defaults or starter templates, could put adopters on the Level~2+ trajectory from the start.

The classifier is also a practitioner tool. It runs on a repository as it stands, requiring no instrumentation, no survey, and no cooperation from the team being measured, so an organization can point it at its portfolio and see which projects have thin AI configuration ahead of a broad agent rollout. Industry maturity frameworks cannot do this: an assessment that depends on organizational self-report cannot be run across a hundred repositories at once.

Two caveats temper this advice. Even Level~2+ repositories experience a significant 27\% increase in cognitive complexity; structured practices reduce quality degradation but do not eliminate it. And we measure the presence of committed artifacts, not the quality of their content or enforcement. Whether the quality or specificity of these artifacts matters is a question for future research.

\subsection{Implications for SE Research}

Practice maturity is a moderator that studies of AI tool impact should measure, ideally as an interaction with treatment rather than a covariate. A study that measures only velocity will find uniform positive effects; one that measures only quality will find a moderate average that hides very different trajectories. Separating the two requires both outcome families and longitudinal designs that show whether effects compound or stabilize over time.
RAMP provides a reusable stratification variable for this, moving analyses past the binary adopted-versus-not comparison.

The instrument also suggests uses beyond stratification. Repository mining studies sample projects on stars, activity, or language, none of which reveal how a project works with AI; RAMP offers a sampling and filtering dimension for mining work that needs to hold AI practice constant or vary it deliberately. A second direction concerns the artifacts. We measure presence, so a placeholder rules file and a carefully maintained one receive the same maturity label; whether specificity, length, or upkeep predicts outcomes is an open research question, and answering it would test the guardrail channel directly. Third, maturity is a repository-level label here, while practice plausibly varies across a repository's subsystems and contributors. Finer-grained measurement could show whether the association we report survives at lower levels where AI is used.

RAMP is a measurement construct, and it can be computed in more than one way. Given the same category definitions, a general-purpose LLM judge matched our embedding-based classifier on the held-out human labels (statistically indistinguishable overall, and somewhat more consistent across categories; Section~\ref{sec:human-validation}), so for most research use we recommend an LLM judge as the simpler option. The embedding pipeline still has one advantage: it runs locally and reproducibly at low cost, which matters for privacy-sensitive corporate repositories that cannot be sent to a hosted model---including those in our own development sample---and for air-gapped or very large-scale settings. Sharing embedding vectors rather than raw files further reduces exposure of proprietary configuration, making cross-organizational benchmarking possible without transmitting source text.

\section{Threats to Validity}
\label{sec:threats}

We organize threats following the four-category framework of Wohlin et al.~\cite{wohlin2012}.

\subsection{Construct Validity}

\textbf{Maturity classification.}
Because the nine semantic categories are defined to map to maturity levels, some cumulative structure is imposed by construction; the Guttman and permutation tests (Section~\ref{sec:validation}) constrain but cannot fully rule out this circularity, and with no Level~4 evidence in the development sample the scale certifies only the Level~2--Level~3 portion, leaving the four-level claim to the mapping-permutation and prevalence-ordering results.
Human annotation validates the labels directly but is bounded by moderate agreement ($\alpha = 0.572$) and sparse support for the rarest categories (Section~\ref{sec:human-validation}).
Because classification relies on committed artifacts, teams that use AI tools without committing configuration are scored Level~1; this attenuates rather than inflates the gradient but makes Level~1 heterogeneous, and it ties maturity partly to tool conventions that mandate committed files (Cramer's V~=~0.483).
The development sample spans 27 volunteering organizations, so a few large ones with consistent internal practices could make the observed structure reflect policy rather than emergent behavior.

\textbf{Treatment and outcomes.}
Higher-maturity repositories may generate more agent pull requests, so the contrast may partly conflate treatment dose with response; the elevated Level~4 duplication estimate (+31.4\%) is consistent with such a dose-response effect, and we do not condition on agent PR volume.
Our outcomes---cognitive complexity, static-analysis warnings, and duplicated line density---are SonarQube proxies rather than direct measures of defects or maintainability, and agent-generated code may differ stylistically in ways that inflate apparent complexity.

\subsection{Internal Validity}

\textbf{Maturity is not randomly assigned.}
Level~2+ repositories may differ from Level~1 on (1)~engineering maturity, (2)~code hygiene, (3)~AI usage intensity, (4)~domain complexity, and (5)~model capability.
The difference-in-differences design absorbs time-invariant confounders and propensity-score matching adjusts observable pre-treatment characteristics, partially addressing (1), (2), and (4); AI usage intensity and model capability are unobserved and cannot be adjusted, and (5) in particular is an alternative explanation of the same pattern.
The gradient is large (roughly 2$\times$ for complexity and 1.7$\times$ for warnings in the agent-first stratum, which holds prior AI exposure constant), but warnings show elevated pre-treatment coefficients and are more weakly identified, so we frame the Study~2 associations as hypothesis-generating rather than causal.

\textbf{Timing and reverse causality.}
Maturity is measured at the end-of-observation snapshot, and most configuration does not predate the agent: of the 273 Level~2+ treated repositories, only 11.0\% committed their first validated artifact before the adoption month, and only 3.8\% within the agent-first stratum where the estimates are identified.
Level~2+ thus largely reflects configuration adopted alongside or after the agent; configuration committed in response to quality problems is a reverse-causality path that biases the gradient downward, so our comparisons likely understate the gap.

\textbf{Identification and stratification.}
RAMP maturity and Agarwal et al.'s agent-first/IDE-first partition overlap: 91 of the 113 IDE-first repositories are Level~2+, so the maturity contrast is identified essentially within the agent-first population (the IDE-first Level~1 stratum, 22 repositories, is too small to estimate separately; cross-tabulation in Appendix~\ref{app:crosstab}).
Because the strata share the original matched control pool (Section~\ref{sec:moderator}), within-stratum covariate balance is not guaranteed, and stratum-specific matching would strengthen future work.

\subsection{External Validity}

Study~1 covers corporate repositories and Study~2 open-source projects; enterprise teams that manage AI practices through policy or internal tooling rather than committed artifacts would be scored Level~1 despite functioning at Level~2+.
RAMP also assigns a single label per repository, whereas AI practice can vary across a repository's subsystems, teams, and contributors---some teams piloting agentic workflows ahead of the rest---so a repository-level label can mask within-repository heterogeneity that finer-grained measurement would need to resolve.
The event study extends six months after adoption and the adoption-dynamics window spans roughly 14 months (Level~1 durations are lower bounds): Level~1 quality paths show little reversion within that horizon, but we cannot rule out later self-correction.

\subsection{Conclusion Validity}

The primary contrast dichotomizes at Level~1 versus Level~2+, and a boundary chosen after inspecting outcomes could overstate the gradient; the four-level, combined Level~1--2, and binary groupings (Section~\ref{sec:robustness}) all locate the dominant gap at the exit from Level~1, so the split follows the empirical structure rather than an outcome-optimized threshold and should be read as descriptive, not as a qualitative threshold at Level~2.
Commit velocity also retains a partial correlation with maturity after volume control ($\rho = 0.269$), so cross-sectional comparisons should adjust for activity.

\section{Conclusion}
\label{sec:conclusion}

We opened with a puzzle: teams report sharply different returns from the same coding agents. Part of the difference, it turns out, is written down. RAMP turns the AI configuration a team commits to version control into a validated four-level maturity scale, and stratifying an existing panel of agent adopters by that scale exposes a clean asymmetry: agents accelerate development at every level, but the quality cost falls almost entirely on repositories that commit no configuration at all. The decisive gap is between committing nothing and committing a few pages of rules and standards.

These associations are observational, and correlated engineering discipline or model capability may explain part of the gradient. What survives that caveat is a way to keep asking the question rigorously: RAMP lets future studies stratify, sample, and filter repositories by how they work with AI rather than by whether they adopted it. The open problems are now about depth---whether the specificity and upkeep of these artifacts predict outcomes, whether the association holds within a repository's subsystems and teams, and whether study designs that fix configuration before adoption can turn this correlation into cause. As agentic development matures, how a team configures its agents may become one of the most consequential things a repository records. It is already one of the most measurable.

\section*{Data Availability}

A replication package is available at: \url{https://doi.org/10.5281/zenodo.21406147}.
The Study~2 outcome data are derived from the publicly released dataset of Agarwal et al.~\cite{agarwal2026}.
The corporate repositories in the Study~1 development sample were contributed under research agreements that prohibit identifying the repositories or organizations; the paper reports their aggregate statistics, and the package includes the full pipeline, tool patterns, and analysis notebooks.

\begin{acks}
We thank Christian K\"astner (Carnegie Mellon University), Simon Obstbaum and Igor Ciobanu (P10Y), and Igor Solomatov (Grid Dynamics) for their valuable feedback and support.
\end{acks}

\balance
\bibliographystyle{ACM-Reference-Format}
\bibliography{software}

\appendix
\counterwithin{table}{section}
\counterwithin{figure}{section}

\section{Human Validation Details}
\label{app:human-validation}

Table~\ref{tab:per-category} reports per-category precision, recall, and F1 against the 175-item human reference of Section~\ref{sec:human-validation}, for the RAMP classifier and the independent LLM-based labeler (Claude Haiku).
Both labelers succeed on the same categories (commands, rules, configuration, agents --- the latter with perfect precision) and fail on the same rare procedural categories (skills, flows), where support is too low for the magnitudes to be estimates.

\begin{table}[h]
\centering
\caption{Per-category precision, recall, and F1 against the human reference (175 items). Rows with support $n < 5$ are anecdotal.}
\label{tab:per-category}
\small
\begin{tabular}{lrrrrr}
\toprule
& & \multicolumn{3}{c}{RAMP classifier} & LLM \\
\cmidrule(lr){3-5} \cmidrule(lr){6-6}
Category & $n$ & P & R & F1 & F1 \\
\midrule
rules (L2)          & 55 & 0.82 & 0.91 & 0.86 & 0.82 \\
configuration (L2)  & 5  & 1.00 & 1.00 & 1.00 & 1.00 \\
architecture (L2)   & 1  & 0.00 & 0.00 & 0.00 & 0.00 \\
code-style (L2)     & 6  & 1.00 & 0.17 & 0.29 & 0.57 \\
agents (L3)         & 19 & 1.00 & 0.63 & 0.77 & 0.77 \\
commands (L3)       & 73 & 0.97 & 0.92 & 0.94 & 0.97 \\
skills (L3)         & 5  & 0.00 & 0.00 & 0.00 & 0.00 \\
flows (L4)          & 3  & 0.00 & 0.00 & 0.00 & 0.00 \\
session-logs (L4)   & 0  & ---  & ---  & ---  & ---  \\
not-artifact        & 8  & 0.30 & 1.00 & 0.46 & 0.58 \\
\bottomrule
\end{tabular}
\end{table}

\begin{figure*}[t]
\centering
\includegraphics[width=0.92\textwidth]{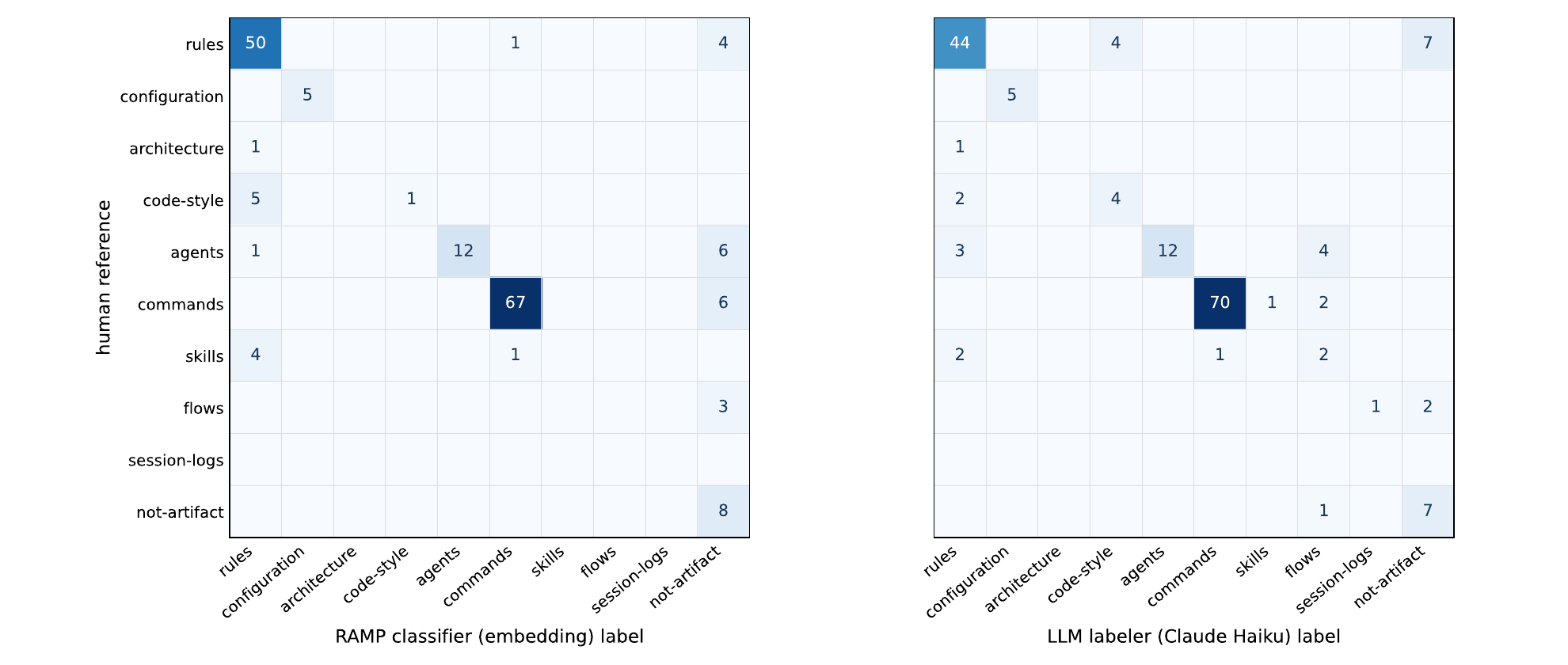}
\caption{Confusion matrices against the 175-item human reference for the RAMP classifier (left) and the LLM labeler (right). Rows are human reference labels, columns machine labels; the ten classes collapse general-documentation and none into not-artifact. Diagonal shares reproduce the file-level accuracies of Table~\ref{tab:human-validation} (81.7\% and 81.1\%).}
\label{fig:confusion}
\Description{Two ten-by-ten confusion matrices side by side, the RAMP classifier on the left and the LLM labeler on the right. Rows are human reference labels and columns machine labels, ordered rules, configuration, architecture, code-style, agents, commands, skills, flows, session-logs, and not-artifact. Counts are printed in the cells and shaded darker for larger values. Both matrices are dominated by the same two diagonal cells, rules and commands: 50 and 67 for the RAMP classifier, 44 and 70 for the LLM labeler. Off-diagonal mass falls in the same places in both: mainly the rare procedural rows, skills and flows, landing in the rules and not-artifact columns. The session-logs row is empty in both.}
\end{figure*}

Figure~\ref{fig:confusion} gives the full confusion matrices underlying Table~\ref{tab:per-category}: both labelers concentrate their errors in the same cells, most visibly the rare procedural categories absorbed into rules or not-artifact.
Because the labeled items cluster within 35 repositories, item-i.i.d.\ intervals are anti-conservative; a repository-cluster bootstrap (10,000 draws) puts file-level accuracy at 63.3--95.6\% for the RAMP classifier and 67.5--90.3\% for the LLM labeler --- the sample supports comparing labelers, not certifying a third digit of accuracy.
Treating each rater in turn as ground truth bounds classifier accuracy at 67.4--84.8\% (LLM: 69.6--75.0\%), inside the human--human agreement range.
On majority-backed references the classifier reaches 93.6\% versus 77.3\% on single-rater references, so its accuracy is highest exactly where the human signal is strongest.
The repository-level exact-match rate of 34/35 has a Wilson 95\% CI of 85--99\% (repository bootstrap: 91.4--100\%).

\section{Additional Classification-Validation Material}
\label{app:validation-material}

\begin{figure}[t]
\centering
\includegraphics[width=\columnwidth]{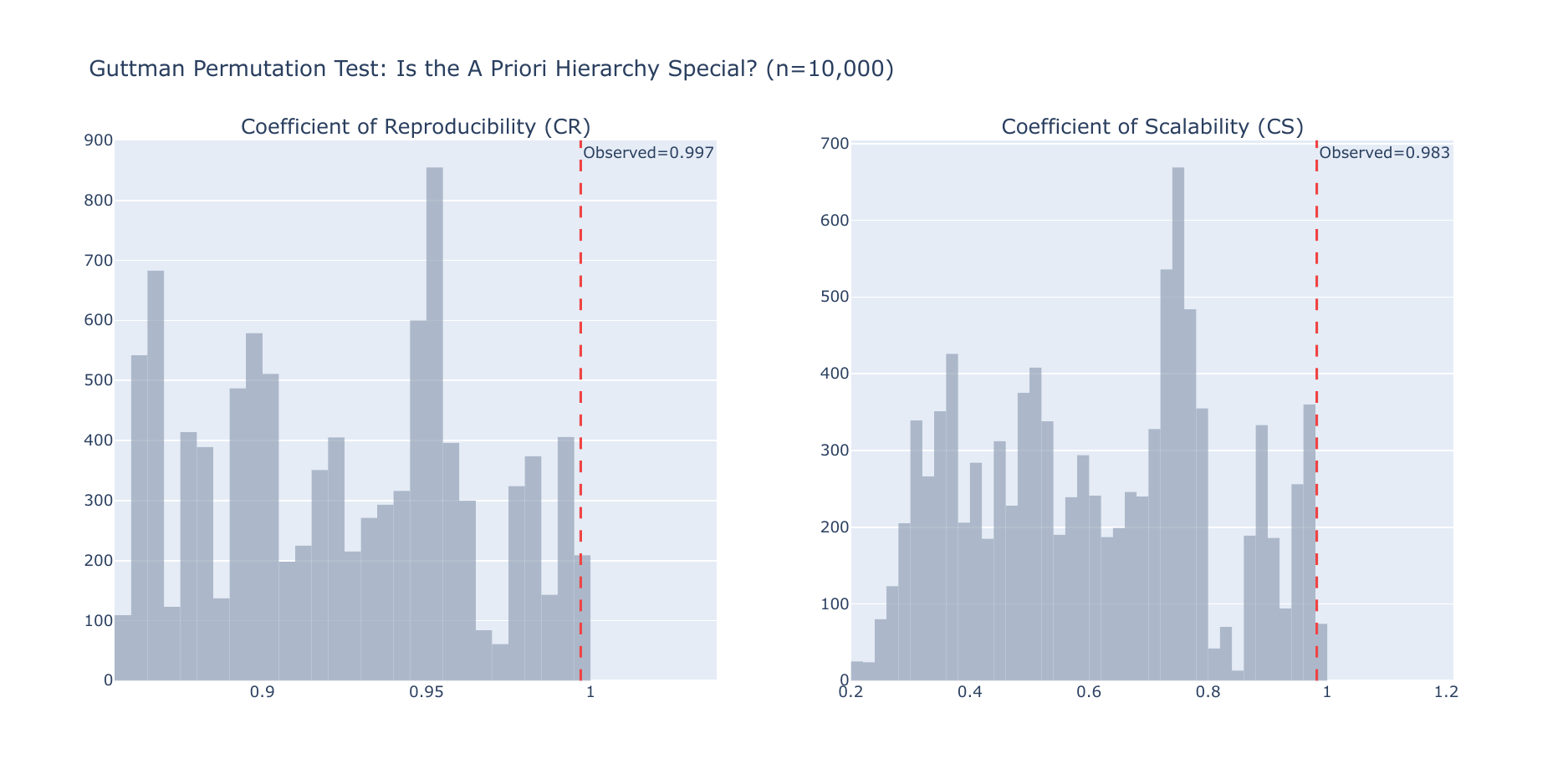}
\caption{Permutation null distributions for the mapping validation (10,000 random category-to-level assignments): the observed CR = 0.997 and CS = 0.983 (dashed) sit at the 99.3rd and 99.5th percentiles.}
\label{fig:guttman-perm}
\Description{Two histograms side by side showing null distributions from 10,000 random category-to-level assignments. The left panel is the coefficient of reproducibility, spanning roughly 0.86 to 1.0; the right panel is the coefficient of scalability, spanning roughly 0.2 to 1.0. Both distributions are irregular and multimodal. A red dashed vertical line marks the observed value in each panel, 0.997 for reproducibility and 0.983 for scalability, and both lines sit in the extreme right tail with only a thin slice of permuted mass beyond them.}
\end{figure}

Figure~\ref{fig:guttman-perm} shows the null distributions behind the mapping-permutation test of Section~\ref{sec:validation}.
The classification-threshold sweep of Section~\ref{sec:validation} is not shown as a figure: all 21 tested thresholds yield the identical classification, so the plot would be a flat line.

\section{Maturity and Prior AI Exposure}
\label{app:crosstab}

Table~\ref{tab:crosstab} cross-tabulates the RAMP maturity level of the 509 classified Study~2 treated repositories against Agarwal et al.'s~\cite{agarwal2026} agent-first versus IDE-first partition.
The maturity contrast is identified essentially within the agent-first population: agent-first repositories split 214 versus 182 between Level~1 and Level~2+, while 91 of the 113 IDE-first repositories are Level~2+, leaving an IDE-first Level~1 stratum of only 22 repositories (Section~\ref{sec:threats}).

\begin{table}[H]
\centering
\caption{RAMP maturity by prior AI exposure for the 509 classified Study~2 treated repositories.}
\label{tab:crosstab}
\small
\begin{tabular}{lrrr}
\toprule
Maturity level & Agent-first & IDE-first & Total \\
\midrule
Level~1 & 214 & 22 & 236 \\
Level~2 & 155 & 55 & 210 \\
Level~3 & 5   & 18 & 23  \\
Level~4 & 22  & 18 & 40  \\
\midrule
Total   & 396 & 113 & 509 \\
\bottomrule
\end{tabular}
\end{table}

\end{document}